\documentclass[aps,showpacs,pra,amssymb,amsmath,twocolumn]{revtex4-2}
\usepackage{graphicx}
\usepackage{amssymb}
\usepackage{amsmath}
\usepackage{amsthm}
\usepackage{bm}
\usepackage{xcolor} 
\usepackage{array}
\usepackage{multirow}
\usepackage{tabularx}
\usepackage{booktabs}
\usepackage{textcomp}
\usepackage{mathtools}
\usepackage{hyperref}

\usepackage{qcircuit}

\usepackage{bbm}
\usepackage{cancel}
\usepackage{comment}
\usepackage{physics}

\begin{document}

\title{Quantum-Enhanced Simulated Annealing Using Rydberg Atoms}
\author{Seokho Jeong, Juyoung Park, Jaewook Ahn}
\address{Department of Physics, KAIST, Daejeon 34141, Republic of Korea}
\date{\today}

\begin{abstract} \noindent
Quantum-classical hybrid algorithms offer a promising strategy for tackling computationally challenging problems, such as the maximum independent set (MIS) problem that plays a crucial role in areas like network design and data analysis. This study experimentally demonstrates that a Rydberg quantum-classical
hybrid algorithm, termed as quantum-enhanced simulated annealing (QESA), provides a computational time advantage over standalone simulated annealing (SA), a classical heuristic optimization method. The performance of QESA is evaluated based on the approximation ratio and the Hamming distance, relative to the graph size. The analysis shows that QESA outperforms standalone SA by leveraging a warm-start input derived from two types of Rydberg atomic array experimental data: quench evolution (QE) (implemented on the Quera Aquila machine) and adiabatic quantum computing (AQC)
(using the experimental dataset archieved in K. Kim et al., Scientific Data {\bf 11}, 111 (2024). Based on these results, an estimate is provided for the maximum graph size that can be handled within a one-day computational time limit on a standard personal computer. These findings suggest that QESA
has the potential to offer a computational advantage over classical methods for solving complex optimization problems efficiently.
\end{abstract}

\maketitle

\section{Introduction}
Quantum computing holds the promise of tackling certain computational problems on a scale beyond the reach of classical counterpart.~{\cite{Feynman_IntJTP_1982,Nielson_Book,Dowling_PhilTrans_2003}}
Various physical systems are currently being actively investigated for their potential in advancing quantum computing.~{\cite{Monroe_RevModPhys_2021,Saffman_RevModPhys_2010,  Wendin_RPP_2017,Arute_Nature_2019,Adams_JOPB_2020}}
Among these, Rydberg atom arrays have recently gained attention  as a promising platform, offering some outstanding features, such as scalability, high qubit connectivity, dynamic reconfigurability, and, most notably, an intrinsic Hamiltonian that naturally maps to the maximum independent set (MIS) problem.~{\cite{Morgado_AVS_QS_2021,Adams_JOPB_2020, Barik_FQST_2024,Saffman_JOPB_2016,Pichler_2018}}

The MIS problem involves finding the largest possible indepedent set (MIS) of vertices for a given graph, $G(V,E)$, where $V$ represents the set of vertices and $E$ represents the set of edges, ensuring that no two vertices in the MIS are connected by an edge in $E$. This problem has numerous practical applications, including network design, social network analysis, and resource allocation.~{\cite{Hale_ProcIEEE_1980, PJWan_IEEE_INFOCOM_2014,Basagni_Telecomm_Syst_2001, Dall_Asta_PRE_2009,KaiSun_PPS_arxiv_2020,Eddy_MIS_Satellite_arxiv_2020}} However, solving large instances of the MIS problem efficiently remains a significant challenge for classical algorithms, as it is generally classified as an non-deterministic polynomial-time (NP)-complete problem in computational complexity theory.~{\cite{Arora_Complexity}} Quantum computing, particularly 
within the framework of adiabatic quantum computing  (AQC) Rydberg atom arrays, is considered as a strategy to tackle this problem efficiently and many applications to MIS graphs have been documented.~{\cite{Loic_PRA_2023, Ebadi_Science_2022, MHK_NP_2022, JSH_PRR_2023, AndyByun_PRXQ_2022, KKH_SD_2024, PJY_PRR_2024, YSong_PRR_2021, AndyByun_AQT_2024, Oliveira_PRXQ_2025}}

A quantum-classical hybrid algorithm has been recently proposed as an approach to overcoming the computational challenges of hard problem instances, such as the MIS problem.~\cite{JWurtz_IEEE_Trans_on_QE_2024,McClean_NJP_2016} In the context relevant to this paper, experimental results from a Rydberg atom array can be utilized as a warm-start input for classical optimization methods, such as the simulated annealing (SA) . This approach, maybe referred to as quantum-enhanced simulated annealing (QESA), can effectively combine the advantages of both classical and quantum methods—leveraging the vast algorithmic resources of classical computing and the massive computational parallelism inherent in quantum many-body systems. As a result, QESA has the potential to reduce computational time compared to the standalone SA method when addressing the MIS problem.~{\cite{JWurtz_IEEE_Trans_on_QE_2024}}

In this paper, we aim to demonstrate the advantage of QESA using experimental results from adiabatic quantum computing (AQC) (with the dataset previously reported~{\cite{KKH_SD_2024}}) and quench evolution (QE) (with newly executed experiments on the Quera Aquila machine~{\cite{Aquila2023}}) with Rydberg-atom arrays, focusing on system scales approximately one hundred atoms. Specifically, we investigate the relationship between QESA's computational time and Hamming distance to show that QESA outperforms SA by utilizing a quantum-experimentally optimized Hamming distance distribution, whereas SA starts with an unoptimized distribution. Building on these results, we examine the scalability of QESA in terms of both the number of atoms and Hamming distance, offering a numerical extrapolation of its potential to efficiently solve larger MIS problems while remaining within the constraints of available classical computational resources.

{
\section{Rydberg atom approach to the Maximum Independent Set Problem} 
Rydberg atom arrays consist of neutral atoms trapped in optical tweezers, where individual atoms are manipulated using laser-induced excitations to Rydberg states.~{\cite{Morgado_AVS_QS_2021,Barik_FQST_2024,Saffman_JOPB_2016,Adams_JOPB_2020}} A key feature of Rydberg atom systems is the Rydberg blockade mechanism. Due to strong dipole-dipole interactions between Rydberg atoms, two neighboring atoms cannot be excited to the Rydberg state simultaneously whenever the atoms are closely displaced than the blockade radius. This inherently enforces the independence condition of the MIS problem, allowing for an efficient mapping of graph problems to physical quantum systems. 

The quantum Hamiltonian governing the Rydberg atom system is expressed (in the unit of $\hbar=1$) as:
\begin{equation} \label{H_q}
\hat{H}_{\mathrm{Quantum}} = \sum_{j \in V} \left( \frac{\Omega(t)}{2} \hat{\sigma}^x_j - \Delta(t) \hat{n}_j \right) + \sum_{(j, k) \in E} \frac{C_6}{r_{jk}^6}
\hat{n}_j \hat{n}_k,
\end{equation}
where $\Delta(t)$ is the laser detuning parameter, $\Omega(t)$ is  the Rabi frequency, $C_6$ is the Van der Waals interaction coefficient, $C_6/r_{jk}^6$, that accounts for the interaction between $j$ and $k$-th atoms distanced by $r_{jk}$, $\hat{\sigma}_j^x$ is Pauli $x$ operator acting on $j$-th atom, and $\hat{n}_j$ is the Rydberg occupation number operator (0 or 1) of $j$-th atom. Proper tuning of $\Omega$ and $\Delta$ allows the system to evolve toward a solution to the MIS problem.~{\cite{YSong_PRR_2021}}

The experimental feasibility of solving the MIS problem with Rydberg atom arrays has been tested,~{\cite{YSong_PRR_2021, Loic_PRA_2023, Ebadi_Science_2022, MHK_NP_2022, AndyByun_PRXQ_2022, JSH_PRR_2023, KKH_SD_2024, PJY_PRR_2024, AndyByun_AQT_2024, Oliveira_PRXQ_2025}} following the formal introduction of the relation between the Rydberg-atom Hamiltonian and the MIS problem.~{\cite{Pichler_2018}}  For example, up to 289 atoms were arranged in a 2D unit-disk graph~{\cite{Ebadi_Science_2022}},  with edges defined by the blockade radius, and by adiabatically sweeping the laser detuning, the system evolved toward approximate MIS solutions, in which the key finding is the superlinear quantum speedup over classical methods like simulated annealing. Other experiments including the implementaion of Rydberg quantum wires for nonlinear graphs,~{\cite{AndyByun_PRXQ_2022}} unit-ball graph embedding using a 3D atom array,~{\cite{MHK_NP_2022}} and non-unit disk graphs,~{\cite{Loic_PRA_2023}} among others.

The MIS problem is an NP-complete class problem, allowing reductions to other NP problems.~{\cite{Arora_Complexity}} Leveraging this NP-completeness of the MIS problem, Rydberg atom graphs have been used to program the Boolean satisfiability (SAT) probem,~{\cite{JSH_PRR_2023}} integer factorization,~{\cite{PJY_PRR_2024}} quadratic unconstrained binary optimization (QUBO),~{\cite{AndyByun_AQT_2024}} and higher order unconstrained binary optimization (HUBO) with hypergraphs.~{\cite{ByunPRA2024}} Furthremore, the maximum weighted independent set (MWIS) problem has been explored with local light shifts.~{\cite{Oliveira_PRXQ_2025}}

Many challenges remain, despote progress, which include scalability, noise,~{\cite{Robicheaux_PRA_2021,Saffman_RevModPhys_2010,BongjuneKim_PRA_2019}} decoherence,~{\cite{Saffman_JOPB_2016}} graph embedding constraints, and benchmarking against classical algorithms.~{\cite{LiuTensorNetwork_SIAMJOnSC_2023}}
While current experiments handle hundreds of atoms,~{\cite{Ebadi_Nature_2021,Ebadi_Science_2022,KKH_SD_2024}} scaling to thousands requires improved coherence times and error mitigation.~{\cite{Mohseni_2025}} Errors from imperfect laser control and spontaneous emission degrade solution quality.~{\cite{Saffman_RevModPhys_2010,Saffman_JOPB_2016,Robicheaux_PRA_2021}} A local detuning strategy is introduced, based on vertex connectivity~{\cite{Yeo_AdvQuanTech_2025}}, to reduce errors and improve optimal solution probability.
Graph embedding remains a challenge,~{\cite{Nguyen_PRXQ_2023}} with 3D Rydberg arrays proposed for complex non-planar graphs.~{\cite{Loic_PRA_2023}} Additionally, rigorous comparisons with classical solvers, such as tensor networks,~{\cite{LiuTensorNetwork_SIAMJOnSC_2023}} are needed. While quantum methods show advantages for certain graph instances, their performance depends on structural properties.~{\cite{Andrist_PRR_2023}} Highly connected graphs remain difficult for quantum solvers, highlighting the need for further research to identify cases where quantum approaches consistently outperform classical methods.
}

\section{Quantum enhanced simulated annealing (QESA)} 
\begin{figure*}[thb!]
    \centering
\includegraphics[width=\textwidth]{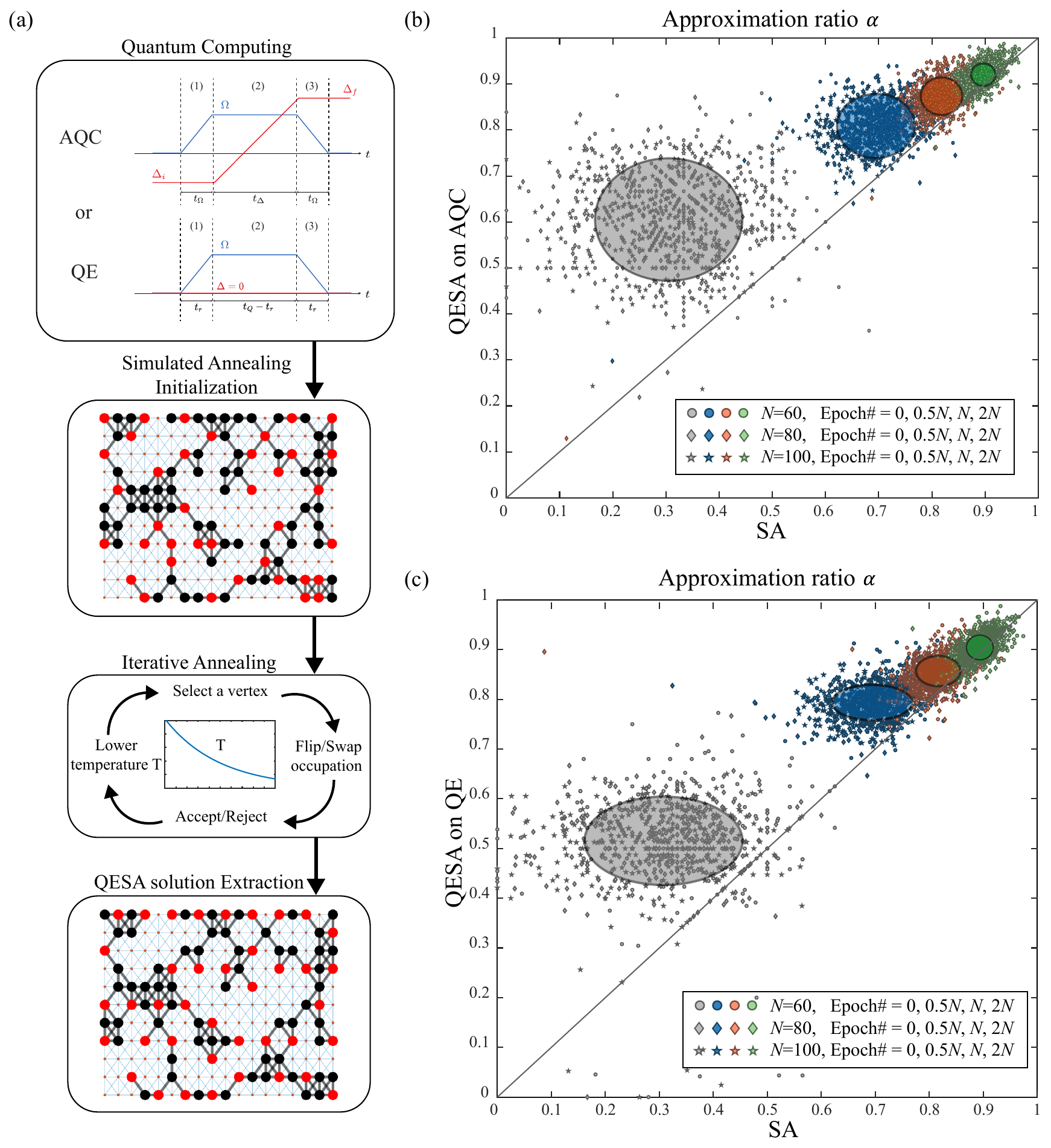}
\caption{(a) The QESA procedure. (b) Comparison of QESA vs. SA for AQC datasets. The approximation ratio ($\alpha$) is plotted for AQC-based QESA (the
$y$-axis) and compared with randomly initialized SA for the chosen graphs (the $x$-axis). A total of 924 graphs (atom arrangements) from the AQC datasets (\#8,\#9, \#10, and \#11)~{\cite{KKH_SD_2024}}, with sizes $N = 60, 80$, and $100$ (number of vertices $N\equiv |V|$), are analyzed and represented by
circular, diamond, and pentagram scatter plots, respectively. Starting from the initial input (gray scatter plot), the SA algorithm progresses through epochs
of $0.5N$ (blue scatter plot), $N$ (orange scatter plot), and $2N$ (green scatter plot). As the number of epochs (Epoch\#) increases, $\alpha$ improves,
indicating better MIS approximations. (c) Comparison of QESA vs. SA for QE datasets. The same analysis is conducted for the above graphs using QE experiments.
In (b) and (c), the size of the ellipses is the standard deviations $\sigma(\alpha)$, representing the spead of $\alpha$.}
\label{Fig1}
\end{figure*}

The procedure of quantum-enhanced simulated annealing (QESA) is illustrated in {Figure~\ref{Fig1}}(a) and consists of four steps. First, in the quantum
computing step, a Rydberg-atom experiment (AQC or QE) is performed to obtain a quantum result for the MIS problem. In the simulated annealing (SA)
initialization step, the quantum result is used to initialize the SA process. Next, the iterative annealing step involves running the SA process iteratively
until a predefined final temperature is reached. Finally, in the QESA solution extraction step, the resulting QESA solution is extracted after the SA process
concludes.

The cost Hamiltonian for the MIS problem is given by:
\begin{equation}\label{H_MIS}
H_{\mathrm{MIS}} = \sum_{j\in V} (-\Delta n_j) + \sum_{(j,k)\in E} Un_j n_k,
\end{equation}
where the occupation number $n_j=0,1$ is a binary variable associated with vertex $j\in V$. The value $n_j = 1$ indicates that vertex $j$ is included in the
independent set, while $n_j = 0$ indicates its exclusion. The constants $\Delta$ and $U$ are positive, with $0 < \Delta < U$. The term $\Delta n_j$ lowers the
energy when $n_j = 1$, promoting the inclusion of vertex $j$ in the independent set. The term $U n_j n_k$ penalizes the inclusion of both vertices $j$ and $k$
in the independent set if they are connected by an edge, thereby increasing the energy to enforce the independence condition. Typically, the parameters are
normalized by setting $\Delta = 1$. For specific graph structures, such as King’s graphs (which represent the legal moves of a king on a chessboard), a value
of $U = 11$ is chosen to strongly penalize the inclusion of adjacent vertices in the independent set.~{\cite{Ebadi_Science_2022}}

For the quantum part of the QESA operation, we utilize either adiabatic quantum computing (AQC) or quench evolution (QE). In AQC experiments, the quantum system is driven quasi-adiabatically, gradually changing $\hat{H}_{\mathrm{Quantum}}$ in Equation~\ref{H_q} over time. As long as the evolution is sufficiently slow, as per the quantum adiabatic theorem, the system remains in the ground state of the time-varying Hamiltonian.~{\cite{Hen_PRA_2015,King_2015,Hauke_RPP_2020,Roenow_Science_2014}} At the end of this evolution, measuring the atoms yields a result corresponding to a low-energy state of $H_{\mathrm{MIS}}$, which provides an approximate solution to the MIS problem. $\Omega(t)$ is chosen typically on the order of $2 \pi \times 1$~MHz and $\Delta(t)$ is changed from $-2 \pi \times 4.0$~MHz to $2 \pi \times 2.0$~MHz~MHz. We use the Rydberg state in $\ket{n=71, S_{1/2}}$. The interaction strength is $U(n=71,r=8.5\,\mathrm{\mu m})=2.7$~MHz for diagonally neighboring atoms and $U(n=71,r=6.0\,\mathrm{\mu m})=21.7$~MHz for laterally neighboring atoms.~{\cite{KKH_SD_2024}}

In QE experiments, the Quera Aquilla machine is driven by a single atom Rabi frequency pulse, with $\Omega=2\pi \times 1$~MHz and under the resonant condition ($\Delta=0$) of the Rabi oscillation for time duration of $t_Q=\pi/[2 \sqrt{\langle \deg(G) \rangle} \cdot \Omega]$, where $\langle \deg(G) \rangle$ is the average degree of vertices in the graph $G$ of atoms. 
This approach is inspired by the graph-degree dependent analysis previously reported.~{\cite{Schiffer2024}} The Rabi frequency rise/fall time $t_r=50$ ns is also taken into account. Unlike AQC, the QE method is expected to be free from algorithmically inherent control errors, allowing it to provide the necessary many-body correlations for solving the MIS problem, even for scalable problem sizes. The $\ket{n=70, S_{1/2}}$ Rydberg state is used where the interaction strength is $U(n=70,r=7.5(1)\,\mathrm{\mu m})=4.8(4)$~MHz for diagonally neighboring atoms and $U(n=70,r=5.3(1)\,\mathrm{\mu m})=39(4)$~MHz for laterally neighboring atoms.

After the first step in the QESA procedure, the atomic states are measured and provided as a ``warm start'' input for the subsequent simulated annealing (SA)
step.~{\cite{JWurtz_IEEE_Trans_on_QE_2024}} The update rules for this SA step follow the ``Rydberg Simulated Annealing" protocol previously reported.~{\cite{Ebadi_Science_2022}} These rules involve three key operatons: First, a free vertex can be added to the independent set. If node $i$ is unoccupied
($n_i=0$) and none of its adjacent vertices are occupied ($n_j = 0$ for all $(i,j) \in E$), vertex $i$ is proposed for inclusion in the set by setting $n_i=1$.
Second, the algorithm can swap the occupation between vertices. If vertex $i$ is occupied ($n_i=1$), its occupation can be swapped with one of its adjacent
vertices. Specifically, for each adjacent vertex $j$, there is a probability of $1/8$ to propose swapping the occupations $n_i, n_j$, leading to the transition $n_i,n_j \rightarrow n_j , n_i$. Lastly, a vertex may be removed from the independent set. If neither of the above applies, vertex $i$ is proposed for removal
by setting $n_i=0$.

Once an update is proposed, the change in the cost Hamiltonian $H_{\mathrm{MIS}}$ associated with the MIS problem is computed. The proposed state is then
either accepted or rejected based on the Metropolis-Hastings criterion~{\cite{Metropolis_1953, Hastings_1970}} at the current temperature $T$. After each
update, $T$ is gradually decreased according to a predefined schedule, and the algorithm proceeds to the next iteration. The iterative annealing step concludes when the temperature reaches a final value of $T=1/\beta = 0.03$,~{\cite{Ebadi_Science_2022}} and the resulting QESA solution is extracted.

\begin{figure*}[th!]
    \centering
\includegraphics[width=\textwidth]{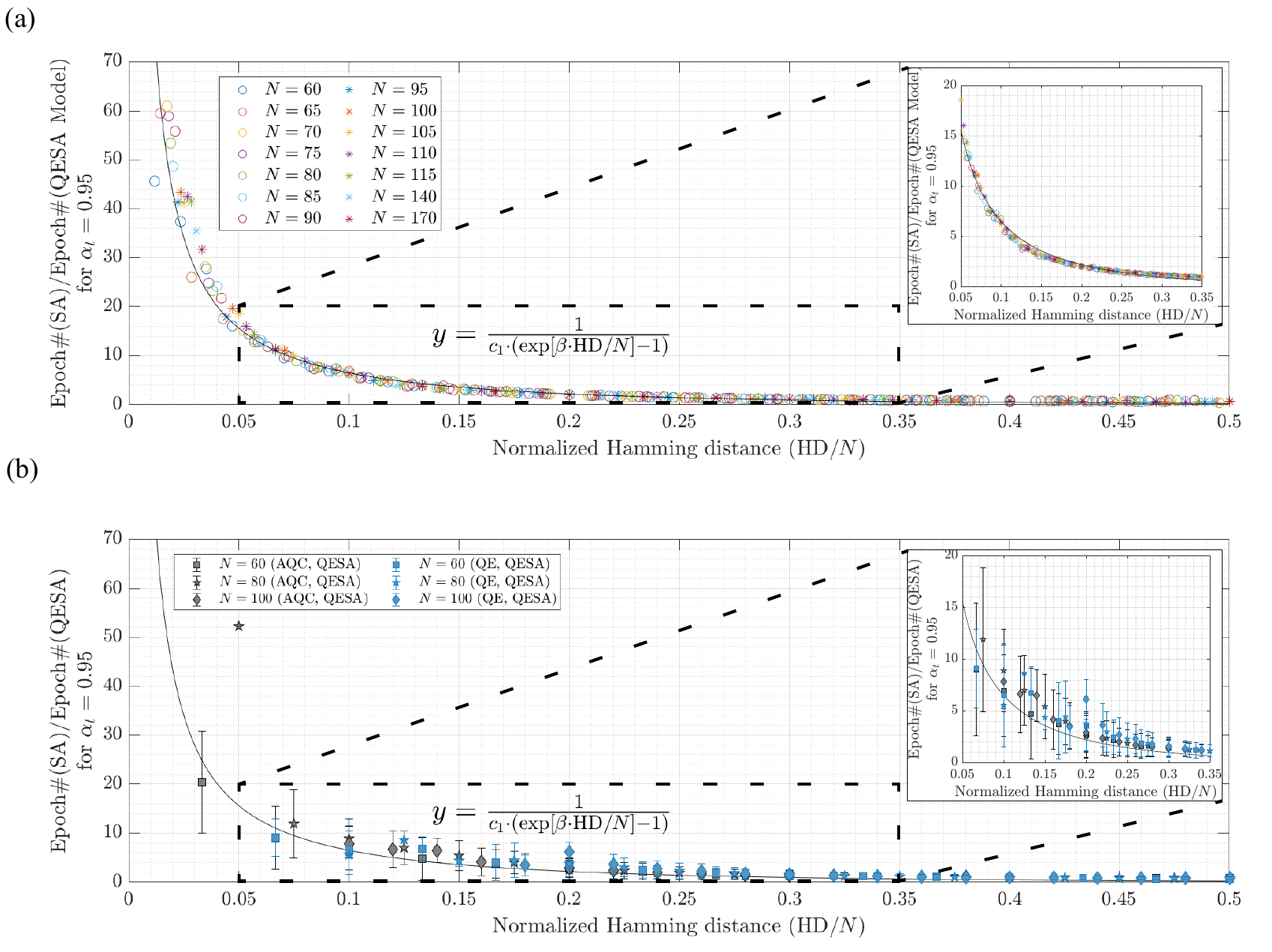}
\caption{The relationship between Epoch\# for $\alpha_t=0.95$ and ${\rm HD}/N$ is depicted in two parts: (a) Epoch\# ratio between SA and QESA models for 1200
graphs with $N=60-115$, 93 graphs with $N=140$, and 154 graphs with $N=170$. The QESA model's monotonic behavior fits well to the black line $y=1/\left[ c_1
\cdot (\exp(\beta \cdot {\rm HD}/N)-1) \right]$, with $c_1=0.1602$, $\beta=6.738$ and an adjusted $R_{\rm adj}^2 = 0.9838$, as detailed in the inset for ${\rm
HD}/N=0.05-0.35$. (b) Epoch\# ratio between SA and QESA for 924 graphs from Figures~\ref{Fig1}(b,c), with error bars representing AQC (black) and QE (blue). }
\label{Fig2}
\end{figure*}

\section{Results}\label{Results} 
To evaluate the performance of QESA and compare it with a corresponding standalone SA, we define a metric, the approximation ratio, $\alpha$, given by
\begin{equation}\label{alpha}
\alpha \equiv \frac{|\{i\in V | n_i = 1\}|-|\{(i,j) \in E | n_i n_j = 1\}|}{|{\rm MIS}|},
\end{equation}
where $|\{i\in V | n_i = 1\}|$ counts the number of vertices in the independent set, $|\{(i,j) \in E | n_i n_j = 1\}|$ counts the number of edges connecting
pairs of occupied vertices that violate the independence condition, and $|{\rm MIS}|$ represents the size of the MIS of the graph. So, $\alpha$ takes into
account two performance criteria: (1) how close the total occupation number is to the MIS solution, and (2) how many edges in the spin configuration violate
the independence condition. It is noted that from the MIS Hamiltonian in Equation~\ref{H_MIS} with $\Delta=1$ and $U=1$, $H_{\rm MIS}(s)\equiv-|\{i\in V |
n_i = 1\}|+|\{(i,j) \in E | n_i n_j = 1\}|$ is an effective enrgy for a spin configuration $s=(n_1, n_2, \cdots)$, which is minimized to $H_{\rm MIS}(s)=-|{\rm MIS}|$ when $s$ is the MIS solution. A value of $\alpha$ close to 1 indicates a high-quality solution that closely approximates the MIS, while lower values
indicate a less accurate approximation.

Figure~\ref{Fig1}(b) shows the evolution of $\alpha$ during the AQC-based QESA method, (using the AQC dataset archieved~{\cite{KKH_SD_2024}}), applied to
various experimental graphs chosen from the datasets, in comparison with pure classical, standalone SA (without AQC data) for the
same graphs. Also, in Figure~\ref{Fig1}(c), the evolution of $\alpha$ during the QE-based QESA method is applied to the same set of graphs as in
Figure~\ref{Fig1}(b), but using the QE experiments implemented on the Quera Aquila platform.~{\cite{Aquila2023}} In both figures, the $y$-axis represents the
performance of QESA, while the $x$-axis shows the results from standalone SA, starting from a random configuration $s$ of $n_i$ with the same total occupation
number $\sum_i n_i$ as the experimental counterpart. The results in Figure~\ref{Fig1}(b,c) show that a substantial proportion—97.5\% in AQC and 91.9\% in QE—of the data points lie above the line $y=x$. Similarly, for ${\rm Epoch\#}/N=0.5$, 1, and 2, the substantial proportions are 98.1\%, 95.5\%, and 87.1\% (based on
AQC results) and 97.0\%, 88.7\%, and 69.9\% (based on QE results). This indicates that QESA consistently outperforms the pure classical SA for the both types of
QESA. These findings highlight the advantage of incorporating Rydberg-atom experimental results as a ``warm start,'' which increases the likelihood of the SA
algorithm converging to higher-quality solutions for the MIS problem compared to starting from a random configuration.

We now focus on evaluating the computation time of QESA, denoted as Epoch\#(QESA), needed to achieve a target approximation ratio (e.g., $\alpha_t = 0.95$) and compare it with the computation time of pure SA, denoted as Epoch\#(SA).
Considering diverse instances which utilize Hamming distance as a robust metric to evaluate and enhance the performance of various heuristic optimization
algorithms (e.g. particle swarm optimization (PSO)~{\cite{Han_IEEE_Trans_on_Sys_2023,Li_IEEE_Trans_on_Rel_2022,Shen_IEEE_Access_2019}}, A$^{*}$ search
algorithms~{\cite{Iordan_J_Adv_Math_CS_2016}} and hypergraph matching problems~{\cite{Kammerdiner_Optim_Lett_2010}}), we identified the Hamming distance, ${\rm HD} \equiv \vert \{ j \in {1, \cdots, N} | s_j \neq t_j \} \vert$ between the initial spin configuration $s$ and the target spin configuration $t$, as a primary relevant variable.

\begin{figure*}[th!]
    \centering
\includegraphics[width=\textwidth]{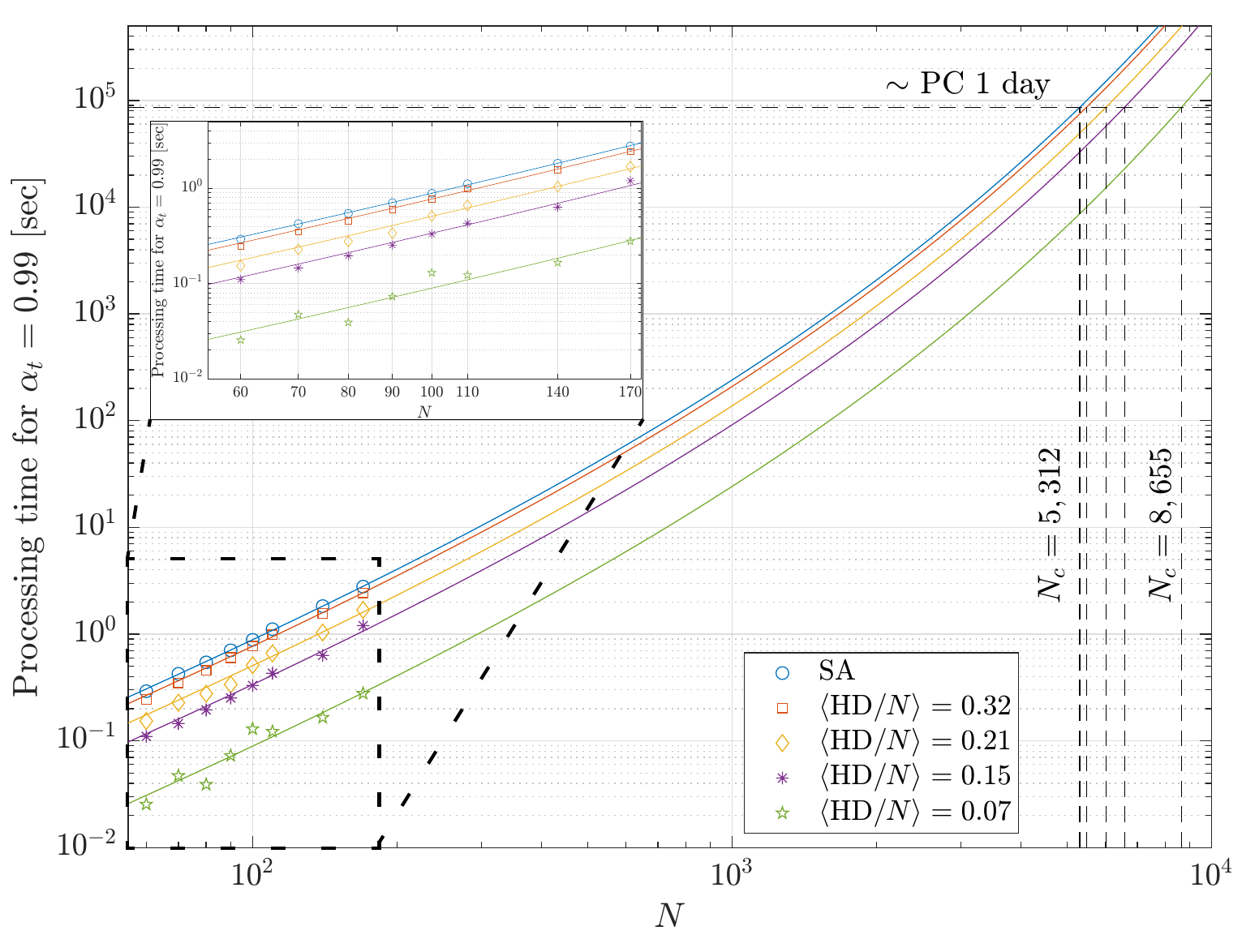}
\caption{ Scalability of graph size $N$ for $\alpha_t=0.99$ within a limited processing time of one day on PC. Each color means SA (blue) case
(blue) and cases with different $\langle {\rm HD}/N \rangle$: 0.32 (orange), 0.21 (yellow) 0.15 (purple) and 0.07 (green).
(Inset: Measured average processing time required to reach $\alpha_t=0.99$ for $N=60$, 70, $\cdots$, 110, 140 and 170.) The required processing time is
estimated as $aN \times b^{\sqrt{N}} \times c N^d$ based on Equation~\ref{SA_Epoch} and the epoch performance ratio in Figure~\ref{Fig2}, where $a_0=5.0508,
b=1.0738$, $c=25.44$ $\mu$s, $d=0.76$ and $a_0/a=1$, 1.15, 1.74, 2.63 and 9.94 for SA and $\langle {\rm HD}/N \rangle=0.32$, 0.21, 0.15 and 0.07 cases,
respectively. From these estimations, the upper bounds $N_c$ of graph size within the one-day limit are predicted to be 5,312 for SA and as 5,484, 6,023,
6,584, and 8,655 for $\langle {\rm HD}/N \rangle=0.32$, 0.21, 0.15 and 0.07 cases, respectively.}
\label{Fig3}
\end{figure*}

As a performance measure of QESA with respect to SA, we consider the ratio between Epoch\# of SA and QESA. In {Figure~\ref{Fig2}}(a,b), we plot the
ratios ${\rm Epoch\#(SA)}/{\rm Epoch\#(QESA \, Model)}$ and\\ ${\rm Epoch\#(SA)}/{\rm Epoch\#(QESA)}$ for $\alpha_t = 0.95$, as functions of ${\rm HD}/N$
(Hamming distance normalized by the graph size $N$), respectively. The former ratio is based on a modeled QESA, detailed below.
Required Epoch\# of simulated annealing is modeled as ${\rm Epoch\#(SA)} = c_1 e^{\beta m^*}$.~{\cite{Hajek_Math_Oper_Res_1988,Bouttier_JMLR_2019}} Since $m^*$can be approximated to ${\rm HD}/N$, we model the ratio as a function $y({\rm HD}/N)$ in Equation~\ref{Epoch_perf_Ratio}:
\begin{equation}
\label{Epoch_perf_Ratio} y({\rm HD}/N) = \frac{1}{c_1 \cdot ( \exp [\beta \cdot {\rm HD}/N]-1 )}, 
\end{equation}
with $c_1=0.1602$, $\beta=6.738$ and an adjusted coefficient of determination, $R_{\rm adj}^2 = 0.9838$, which explains over 98\% of the variability in the
data.

The modeled QESA results in Figure~\ref{Fig2}(a) are obtained through the following procedure: First, we sample SA spin configurations that reached
$\alpha_t=0.95$ from a set of 1200 experimental graphs~{\cite{KKH_SD_2024}} with $N=60-115$ in Exp\#10, 93 graphs with $N=140$, and 154 graphs with $N=170$, inorder to determine their epoch differences for initial points $\alpha_i=0.85, 0.88, 0.91$ across $N=60, 65, 70, \cdots, 115, 140, 170$. Then, we sort these
epoch differences by ${\rm HD}$ and compute the epoch performance ratio, denoted as\\ ${\rm Epoch\#(SA)}$/${\rm Epoch\#(QESA \, Model)}$, which represents the
ratio of epochs for SA to those for QESA for each $\alpha_i$, based on the sorted epoch differences.

In Figure~\ref{Fig2}(b), the ratios ${\rm Epoch\#(SA)}/{\rm Epoch\#(QESA)}$ for AQC and QE (black and blue error bars, respectively) closely follow the black
fitting line from the QESA model. The zoomed-in plots for ${\rm HD}/N=0.05-0.35$ in the insets of Figures~\ref{Fig2}(a,b) show that
Equation~\ref{Epoch_perf_Ratio} characterizes the average epoch performance ratio. This result indicates that QESA's computation time is reduced with a
shorter, quantum-mechanically warm-started ${\rm HD}/N$ compared to the randomly started SA, giving QESA an advantage.

\section{Discussion}\label{Discussion} 
We now discuss the scaling challenges related to the QESA approach. Since QESA is a subset of SA that begins at a shorter Hamming distance, we analyze its
scaling behavior in comparison to that of SA. Simulated annealing (SA) is a probabilistic and iterative optimization algorithm designed to find the optimal
value of a cost function, such as $H_{\rm MIS}$ in our case. The probability of successfully finding the MIS is expressed as $P_{\rm MIS,SA} = 1-\exp
(-a\mathcal{HP}^{-1}/N\cdot {\rm Epoch\#(SA)})$, where $\mathcal{HP}$ is the hardness parameter of the given graph; for King's graphs, it is examined as $\ln
\mathcal{HP}\sim \sqrt{N}$.~{\cite{Ebadi_Science_2022}} Consequently, the epoch-to-solution (ETS)~{\cite{Lidar_PRX_2018}} for the SA algorithm, denoted by
${\rm ETS}_{\rm MIS,SA}$, is given by
\begin{equation} \label{SA_Epoch}
{\rm ETS}_{\rm MIS,SA} = a  N \times \mathcal{HP} \sim a N e^{b\sqrt{N}},
\end{equation}
where $a$ is a coefficient associated with the normalized Hamming distance ${\rm HD}/N$ (inversely proportional to the epoch performance ratio in
Equation~\ref{Epoch_perf_Ratio}), and $b$ is a parameter related to $\mathcal{HP}$ of the graph. This exponential scaling with $\sqrt{N}$ implies that as $N$ increases, the number of iterations required for the algorithm to converge grows significantly.

We assume that the average processing time per epoch, $t_{\rm step}$, scales as a function of $N$:
\begin{equation} \label{t_step}
t_{\rm step} = cN^d,
\end{equation}
where the parameters $c$ and $d$ are determined by the specifications of the PC.
Consequently, the total processing time for target $\alpha_t$, denoted as $t_{\rm processing}$, is obtained as the product of Equation~\ref{SA_Epoch} and
\ref{t_step}:
\begin{eqnarray} \label{t_processing}
t_{\rm processing} &\equiv& t_{\rm step} \times {\rm ETS}_{\rm MIS,SA} \nonumber \\
&=&aN \times b^{\sqrt{N}} \times  c N^d.
\end{eqnarray}
The scalability of $t_{\rm processing}$ is drawn in the inset in {Figure~\ref{Fig3}}, as ${\rm HD}/N$ varies and $N$ increases for $\alpha_t =
0.99$. For $N=60$ to 170, $t_{\rm processing}$ shows performance gaps of 1.15, 1.75, 2.63, and 9.94 times compared to the SA case, with $\langle {\rm
HD}/N\rangle = 0.32$, 0.21, 0.15, and 0.07, respectively. In Figure~\ref{Fig3}, $a$, related to ${\rm HD}/N$, is fitted as $a=5.05$, 4.39, 2.89, 1.92, 0.51 forthe SA case and $\langle {\rm HD}/N\rangle = 0.32$, 0.21, 0.15, and 0.07. $b$, related to $\mathcal{HP}$, is fitted as 1.0738, with $R^2_{adj}$ values of
0.9981, 0.9929, 0.9758, 0.9625 and 0.8276 for SA case and $\langle {\rm HD}/N\rangle = 0.32$, 0.21, 0.15, and 0.07, respectively. $c$ and $d$ are related to
the PC computing environment and fitted with values of 25.44~$\mu$s and 0.76, respectively, yielding $R^2_{adj}$ values of 0.9826. The PC operational environment is MATLAB R2023a with an AMD Ryzen Threadripper 3960X 24-Core processor, NVIDIA GeForce RTX 3080 GPU, and 64.0 GB of RAM. It is assumed that the RAM does not significantly influence the extrapolation.

We now attempt to estimate the maximum graph size $N_c$ that can be processed when $t_{\rm processing}$ is constrained by a finite PC operation time, such as
within 1 day. Extrapolating the numerically fitted scaling as Figure~\ref{Fig3}, $N_c=$5,312 is expected for SA alone. However, for the QESA approach, $N_c$
could increase to 5,484, 6,023, 6,584 and 8,655, for graphs having initial Hamming distances, ${\rm HD}/N=0.32$, 0.21, 0.15 and 0.07, respectively. This
suggests that QESA is capable of handling larger problem instances than the SA algorithm alone.

Furthermore, QESA includes an occupation swap process, which corresponds to spin exchanges in the Rydberg SA process. In contrast, this occupation swap process is not present in the post-processing method~{\cite{Ebadi_Science_2022}} previously reported. As a result, the distribution of $\alpha$ in QESA tends to be more concentrated within a sharper interval and exhibits a higher mean value compared to the corresponding distributions obtained from the post-processing method. This suggests that QESA achieves a solution closer to the optimal value. 

The dataset is available on {{\it figshare} data repository}
(\href{https://doi.org/10.6084/m9.figshare.c.7639259.v1}{https://doi.org/10.6084/m9.figshare.c.7639259.v1}).~{\cite{Dataset_JSH_2025}} The dataset consists of the codes
(\href{https://doi.org/10.6084/m9.figshare.28260860.v1}{https://doi.org/10.6084/m9.figshare.28260860.v1}), data
(\href{https://doi.org/10.6084/m9.figshare.28254581.v1}{https://doi.org/10.6084/m9.figshare.28254581.v1}) and figures
(\href{https://doi.org/10.6084/m9.figshare.28256084.v1}{https://doi.org/10.6084/m9.figshare\\.28256084.v1}) used for this paper.

\section{Conclusion}
This paper has presented a data-driven case study highlighting the computational time advantage of QESA over SA, utilizing AQC and QE
experimental datasets from Rydberg atom arrays with approximately 100 atoms. By analyzing SA’s epoch time dependence on Hamming distance, we have shown that
QESA outperforms SA by utilizing a quantum-optimized Hamming distance distribution, unlike SA’s unoptimized counterpart. Consequently, QESA, an SA variant witha quantum-prepared ``warm start,'' is expected to handle larger graph sizes—up to 8,655 vertices compared to 5,312 for standalone SA within a one-day PC
runtime—allowing for the solution of larger MIS problem instances while staying within classical computational constraints.

\begin{acknowledgements} \noindent
This research was supported by National Research Foundation of Korea (NRF)
grant No. RS-2024-00340652, funded by Korea government (MSIT).
\end{acknowledgements}

\end{document}